\documentclass[11pt, oneside]{article}   	
\usepackage{graphicx}							
\usepackage{amsmath,amsthm,amssymb}

\theoremstyle{definition}

\theoremstyle{remark}

\title{Spectral properties of nonlinear Schr{\"o}dinger equation on a ring }

\author{
Takaaki Nakamura, Taksu Cheon \\
{\it Laboratory of Physics, Kochi University of Technology},\\
{\it Tosa Yamada, Kochi 782-8502, Japan}
}

\date{\today}

\begin{document}
\maketitle
\begin{abstract}
The stationary states of nonlinear Schr{\"o}dinger equation on a ring with a defect is numerically analyzed.  Unconventional connection conditions are imposed on the point defect, and it is shown that the system displays energy level crossings and level shifts and associated quantum holonomies in the space of system parameters, just as in the corresponding linear system.  In the space of nonlinearity parameter, on the other hand, the degeneracy occurs on a line, excluding the possibility of any anholonomies.  In contrast to the linear case, existence of exotic phenomena such as disappearance of energy level and foam-like structure are confirmed.
%
\end{abstract}

\section{Introduction}
Gross and Pitaevskii found a striking fact that many boson system with the short-range interaction in its dilute limit is effectively described under Hartree-Fock approximation by a single wave function that satisfies an wave equation similar to the Schr{\"o}dinger equation with a critical difference of nonlinearity \cite{GR61, PI61}.  It is natural to ask a question how different the properties of nonlinear Schr{\"o}dinger waves are from its linear counterparts.  Such question should be most easily asked in its simplest setting in one-dimension.  In light of experimental progress in recent years on Bose-Einstein condensate in one dimensional devices, such study has obtained unexpected urgency \cite{EXP}.  The nonlinear Schr{\"o}dinger equation appears not only in model equations describing Bose-Einstein condensate, but also in various other physical systems such as vortex motion \cite{VOF}, precession of spin \cite{SPI}, pulse wave in optical fiber \cite{OPF}.
There is a great significance in examining the physical and mathematical properties of nonlinear Schr{\"o}dinger equation in detail.

It is known that the inverse scattering method, or the method of Hirota lead to an exact $n$-soliton solution of the nonlinear Schr{\"o}dinger equation\cite{ISM}, in which there are infinite numbers of conserved quantities.
Moreover, Cazenave and others have found critical results on the stability of the stationary wave of the nonlinear Schr{\"o}dinger equation on a one-dimensional infinite line \cite{STA, STA2, DLT, TU03}.

Although the solution of nonlinear Schr{\"o}dinger equation in infinite line has been known for quite some time, its study on {\it graphs} made up of one-dimensional finite lines and vertices has not started until recently.  Some interesting properties are now being uncovered \cite{TU08, AK96, TH03, TN08, ST94}.  Most studies concentrates on different graph topologies.  The implication of {\it generalized connection condition}, which is known to bring quite a richness to linear Schr{\"o}dinger waves on graphs \cite{TC12, TC13}, has been mostly overlooked until now, with a few exception \cite{DLTP, SP18}.

In this article, we place the nonlinear wave described by the Schr{\"o}dinger equation with cubic nonlinearity on a ring with a single defect that is described by the connection condition specified by two parameters, Fulop-Tsutsui scaling factor and the delta potential strength.  The Fulop-Tsutsui scaling factor connects the wave function at the vertex with the self-adjoint condition and has an interesting property of being scale invariant \cite{FT,FTG,FT2}.   Our choice of connection condition should be considered as a starting trial for the exploration of complete self-adjoint connection condition, which is described by four parameters.
It is found that this nonlinear system displays the quantum holonomies, both Berry phase \cite{BE84} and exotic types \cite{CT09}, 
just like its linear counterpart.
It is also found that the nonlinear system has its unique feature of possessing foam-like energy surface in parameter space.

\section{Nonlinear Schr{\"o}dinger equation on a ring with a defect}
We consider the cubic nonlinear Schr{\"o}dinger equation
\begin{eqnarray}
\label{ee1}
i \dot{\Psi}(x,t) = - \Psi''(x,t) + g \Psi^*(x,t)\Psi(x,t)^2 ,
\end{eqnarray}
on a ring with circumference $2\pi$, namely, a finite line
\begin{eqnarray}
\label{ee2}
x \in [0, L)
\end{eqnarray}
with both ends of the line, $x=0$ and $x=L$ identified.  The dot and prime signify temporal and spatial derivatives, respectively.  At this identified end point, we impose the connection condition
\begin{eqnarray}
\label{ee3}
\Psi'(0,t) - t \Psi'(L,t) = v \Psi(0,t)
\nonumber \\
t \Psi(0,t) - \Psi(L,t) = 0 ,
\end{eqnarray}
in which $v$ and $t$ are real numbers \cite{FT}.
This amounts to considering a ring with a point defect located at $x=0$ (or identically, $x=L$).
This is a subset of the connection condition that guarantees the conservation of the wave function norm, 
\begin{eqnarray}
\label{ee4}
\mu = \int_0^{L} dx \Psi^*(x,t) \Psi(x,t),
\end{eqnarray}
which sometime is also termed the {\it mass}, in the literature.
With an Ansatz $\Psi(x,t) = \psi(x) e^{-i E t}$ where $E$ is a real number and $\psi$ a real function, we have an eigenvalue equation
\begin{eqnarray}
\label{ee11}
-\psi''(x) + g \psi^3(x) = E \psi(x)
\end{eqnarray}
The connection condition (\ref{ee3}) is expressible, in terms of time independent wave function $\psi(x)$ as
\begin{eqnarray}
\label{ee13}
\psi'(0) - t \psi'(L) = v \psi(0)
\nonumber \\
t \psi(0) - \psi(L) = 0
\end{eqnarray}
and the mass is also expressed as
\begin{eqnarray}
\label{ee14}
\mu = \int_0^{L} dx \psi^*(x) \psi(x)
\end{eqnarray}
With the rescaling $\psi=\sqrt{\mu}{\tilde \psi}$, we have
\begin{eqnarray}
\label{ee15}
-{\tilde \psi}''(x) + {\tilde g} {\tilde \psi}^3(x) = E {\tilde \psi}(x)
\end{eqnarray}
with $\int_0^{L} dx{\tilde \psi}^*(x) {\tilde \psi}(x) =1$ and rescaled coupling constant ${\tilde g} = g \mu$.  Therefore we loose no generality by setting the mass to be one,  $\mu=1$.  We shall adopt this convention in the rest of this paper.

The eigenvalue problem (\ref{ee11}) with the mass constraint $\mu = 1$  and the connection condition (\ref{ee13}) at the defect yield eigenvalues $E_n$ and $\psi_n(x)$ ($n=1, 2, ...$) which are the functions of the connection parameter $t$, $v$, and the coupling $g$.  How the energy levels $E_n(t, v, g)$ and the eigenfunctions  $\psi_n(x; t, v, g)$ behave as in the parameter space $(t, v, g)$ is the matter of our interest.  Identifying the location and the mode of level degeneracies in the parameter space is the key to our study.

\section{Solutions in terms of elliptic integrals}

The eigenvalue equation (\ref{ee11}) is fully integrable and its solution can be expressed in terms of elliptic functions $sn$ , $cn$ and $dn$.  
For $g>0$, $E>0$  the solution takes the form
\begin{equation}
\label{ee21}
\displaystyle 
\psi (x)=k_-sn[\sqrt{\frac{g}{2}}k_+(x-x_0),\frac{k_-^2}{k_+^2}]e^{i \eta_0} 
\qquad  
(0\leqq c ) ,
\end{equation}
or
\begin{equation}
\label{ee22}
\displaystyle 
\psi (x)=\frac{k_+}{sn[\displaystyle
 \sqrt{\frac{g}{2}}k_+(x-x_0),\frac{k_-^2}{k_+^2}]}e^{i \eta_0} 
 \qquad  
(0\leqq c \leqq \frac{E^2}{4g}) ,
\end{equation}
and 
\begin{equation}
\label{ee23}
\displaystyle 
\psi (x)=\frac{k_+}{cn[\displaystyle
 \sqrt{\frac{g(k_+^2-k_-^2)}{2}}(x-x_0),-\frac{k_-^2}{k_+^2-k_-^2}]}e^{i \eta_0}  
\qquad  
(c \leqq 0)  .
\end{equation}
Here, $k_\pm$ are defined by
\begin{eqnarray}
\label{ee31}
k_\pm = \sqrt{\frac{\mu}{g}} \sqrt{1 \pm \sqrt{1-\frac{4gc}{\mu^2}} } ,
\end{eqnarray}
and two real numbers $c$ and $x_0$ are the constants of integral.
%
For $g>0$, $E<0$, it is given by
\begin{equation}
\label{ee24}
\displaystyle 
\psi (x)=k_-sn[\sqrt{\frac{g}{2}}k_+(x-x_0),\frac{k_-^2}{k_+^2}]e^{i \eta_0} 
\qquad  
(\frac{E^2}{4g} \leqq c ) ,
\end{equation}
%
\begin{equation}
\label{ee25}
\displaystyle 
\psi (x)=\frac{k_-}{sn[\displaystyle
 \sqrt{\frac{g}{2}}k_-(x-x_0),\frac{k_+^2}{k_-^2}]}e^{i \eta_0} 
 \qquad
 (0 \leqq c \leqq \frac{E^2}{4g}) ,
\end{equation}
and by
\begin{equation}
\label{ee26}
\displaystyle 
\psi (x)=\frac{k_+}{cn[\displaystyle
 \sqrt{\frac{g(k_+^2-k_-^2)}{2}}(x-x_0),-\frac{k_-^2}{k_+^2-k_-^2}]}e^{i \eta_0}
 \qquad  
(c \leqq 0) .
\end{equation}
%
For $g<0$, $E>0$, it takes the form
\begin{equation}
\label{ee27}
\displaystyle 
\displaystyle 
\psi (x)=k_+sn[\sqrt{\frac{g}{2}}k_-(x-x_0),\frac{k_+^2}{k_-^2}]e^{i \eta_0} \hspace{3mm}  (0 \leqq c) ,
\end{equation}
and no solution exist for $c > 0$.
Finally, for $g<0$, $E<0$, it is given by
\begin{eqnarray}
\label{ee28}
\psi(x) = k_+cn[\sqrt{\frac{-g(k_+^2-k_-^2)}{2}}(x-x_0),\frac{k_+^2}{k_+^2-k_-^2}]e^{i \eta_0} 
\qquad  
(0 \leqq c)
\end{eqnarray}
and by
\begin{eqnarray}
\label{ee29}
\psi(x) = k_+dn[\sqrt{\frac{-g}{2}}k_+(x-x_0),\frac{k_+^2-k_-^2}{k_+^2}]e^{i \eta_0} 
\qquad  
(\frac{E^2}{4g} \leqq c \leqq 0) ,
\end{eqnarray}
and no solution exist for $c < \frac{E^2}{4g}$.
Finally, for $g<0$, $E=0$, we have
\begin{eqnarray}
\label{ee30}
\psi(x) =\sqrt{\frac{2}{g}}\frac{1}{x-x_0}e^{i \eta_0} .
\end{eqnarray}
%
%

By imposing the connection condition (\ref{ee13}), and the normalization condition $\mu = 1$ (\ref{ee14}) as self consistent constraints, we can numerically obtain allowable sets of $x_0$, $c$, and $E$ simultaneously, for a given numbers of $t$, $v$, and $g$.

The linear limit $g=0$ is solvable in terms of trigonometric functions
\begin{eqnarray}
\label{ee32}
\psi(x) = \sin{k (x-x_0)}
\end{eqnarray}
with the integral constant
\begin{eqnarray}
\label{ee33}
x_0 = -\frac{1}{k} \arctan{\frac{\sin{k L}}{t-\cos{k L}}} ,
\end{eqnarray}
and with $k=\sqrt{E}$ which is given as the solution of
\begin{eqnarray}
\label{ee34}
t \cos{k L} = 1 \pm \sqrt{ 1 - \cos{k L} \left(\cos{k L} + \frac{v}{k}\sin{k L} \right) } .
\end{eqnarray}
From (\ref{ee31}), we immediately obtain the function $t=t(k, v)$, from which $E=E(t,v)$ is obtainable. 
%

\section{Energy numerics}

%
\subsection{Energy levels as functions of nonlinearity parameter}

With the method detailed in the previous section, we evaluate the energy eigenvalues numerically with Newton's method.
Figure 1 shows the energy eigenvalues as functions of nonlinearity parameter $g$ with a defect parameters $v=0$ and $t=1$, namely, the free connection condition. 
All energy eigenvalues are seen to increase monotonically with $g$.  The level intersection and branching, characteristic to nonlinear system, are observed. 

There are two types of energy lines in the figure; first is the ones obtained from  the equations (\ref{ee21}), (\ref{ee24}), (\ref{ee27}) and (\ref{ee28}), and the other is ones obtained from the equations (\ref{ee22}), (\ref{ee23}), (\ref{ee25}), (\ref{ee26}) and (\ref{ee29}). The straight line passing through the origin in Figure 1 belong to the first type, while the energy level branching from the fisrt type in the negative side of $g$  belong to the second type.  
%

Since the energy level grows with nonlinearity, the second harmonic of the wave function and the third harmonic satisfy the connection conditions (\ref{ee13}) and the mass constraint $\mu=1$, so for the arbitrary nonlinearity and energy eigenvalue pair ($g,E$) in Figure 1, the pair ($n^2g, n^2E$) also satisfies the conditions (\ref{ee13}) and $\mu=1$.

\begin{figure}[htbp]
\begin{center}
\includegraphics[width=4.5cm]{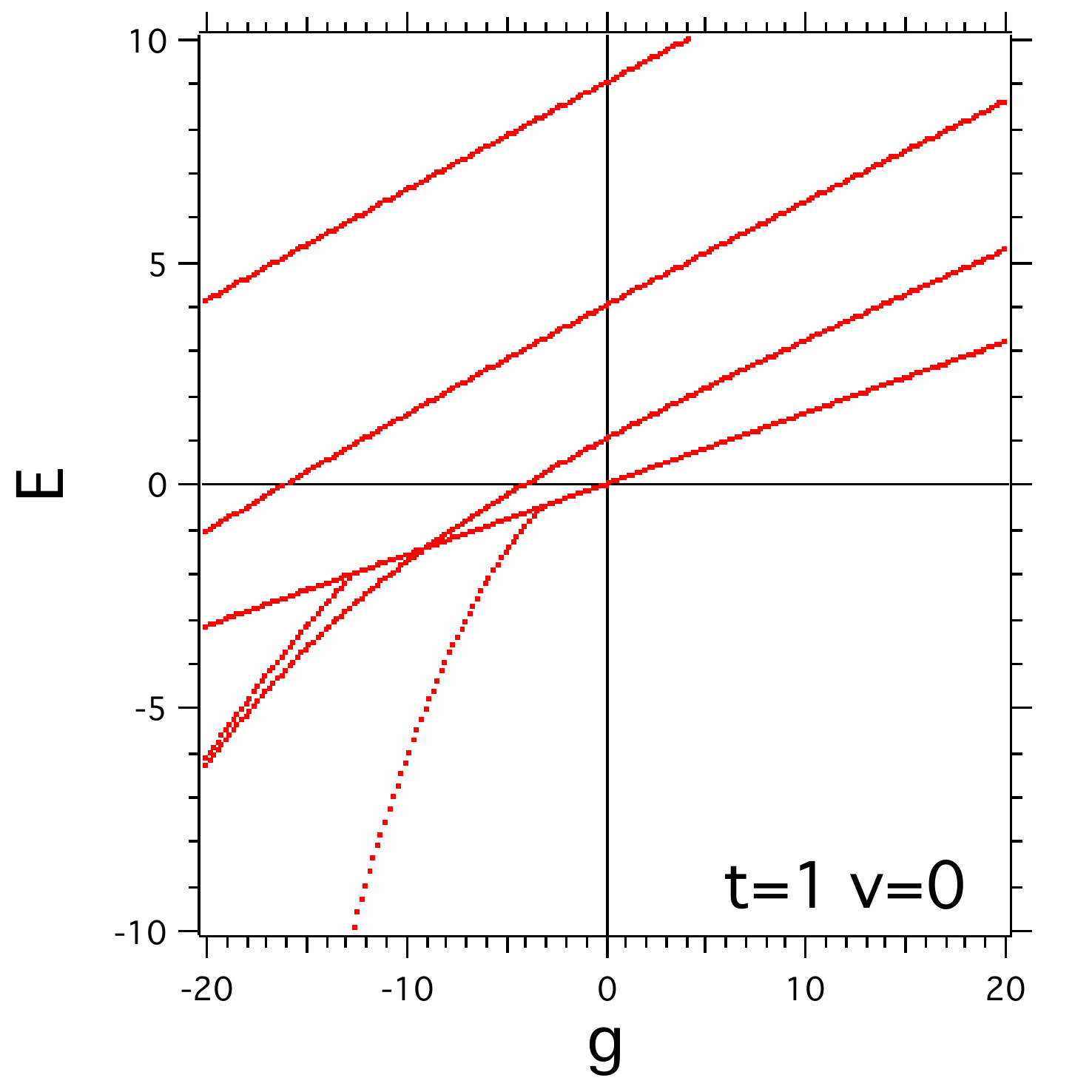}
\end{center}
\caption{The energy levels for free connection condition $t=1$, $v=0$.}
\end{figure}

%
\subsection{Energy levels of linear system as functions of connection condition parameters}

We now calculate the energy eigenvalue $E$ as a function of connection parameters $t$ or $v$, with fixed nonlinearity parameter $g$. 
We start by looking at the linear case $g=0$, as the reference to be compared to the nonlinear cases.
The results are shown in Figure 2, in which the vertical axis represents the scaled energy eigenvalue ${\rm sgn}(E) \sqrt{E}$ and the horizontal axis represents the scaled connection parameter $2\arctan{(t)}$.  The $\delta$-strength parameter $v$ is fixed to be $v=-1$, $0$ and $1$ for the left, middle, and right graphs, respectively. 
With the connection condition $v=0$,  the level crossing occurs at $t = 1$ and $1$.  With other values for $v$, the level repulsion occurs near $t=1$ and $-1$.
%
\begin{figure}[htbp]
\begin{center}
\includegraphics[width=4.1cm]{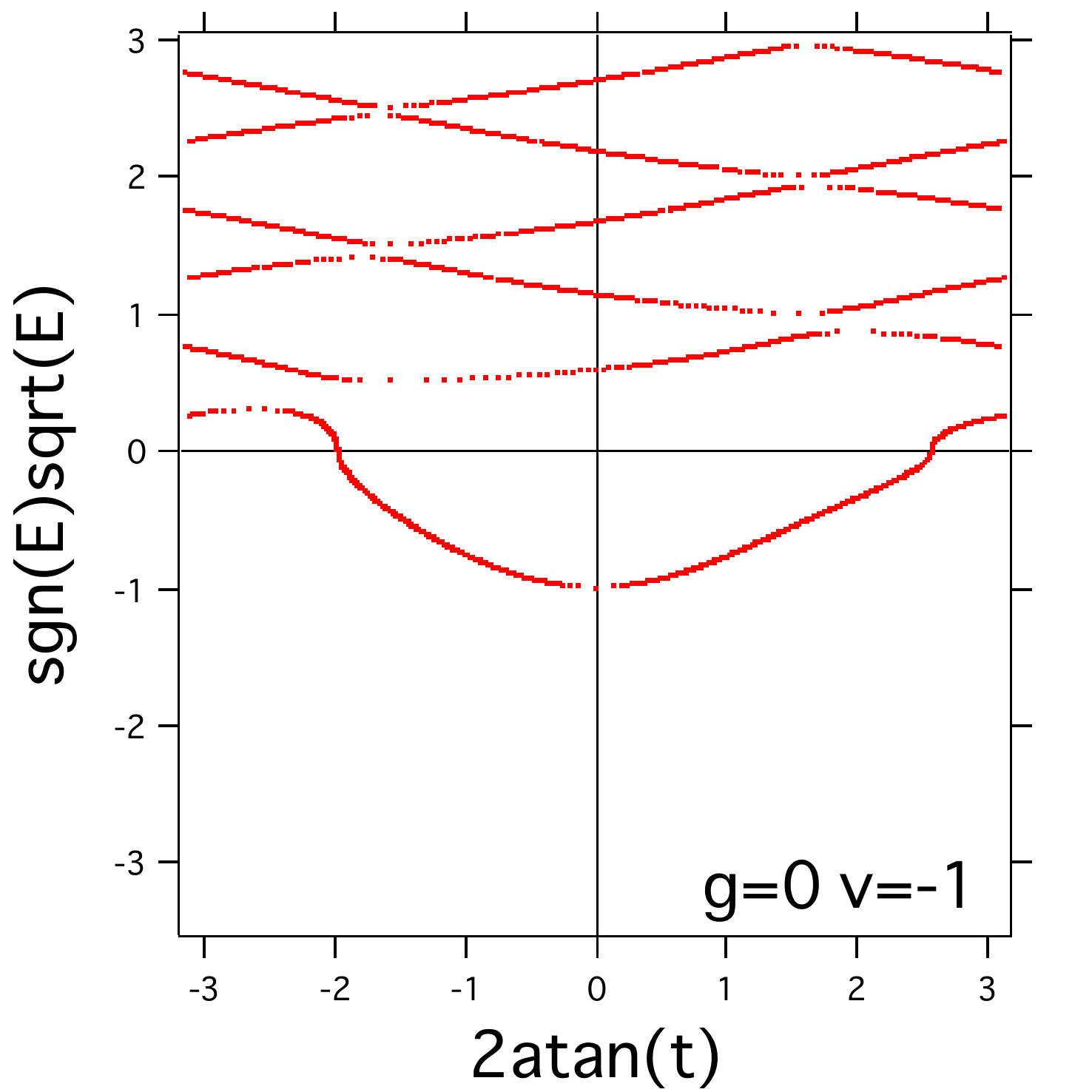}
\includegraphics[width=4.1cm]{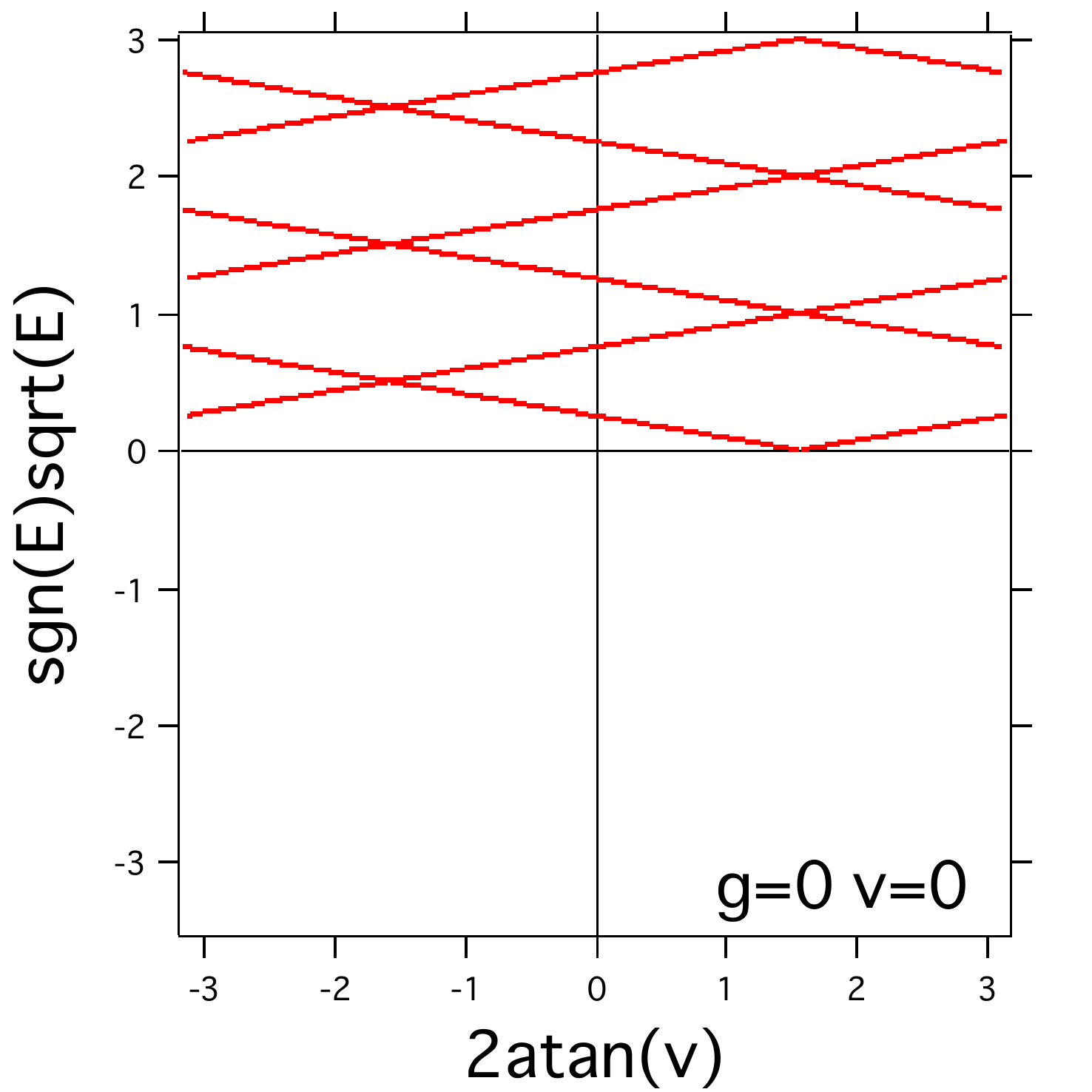}
\includegraphics[width=4.1cm]{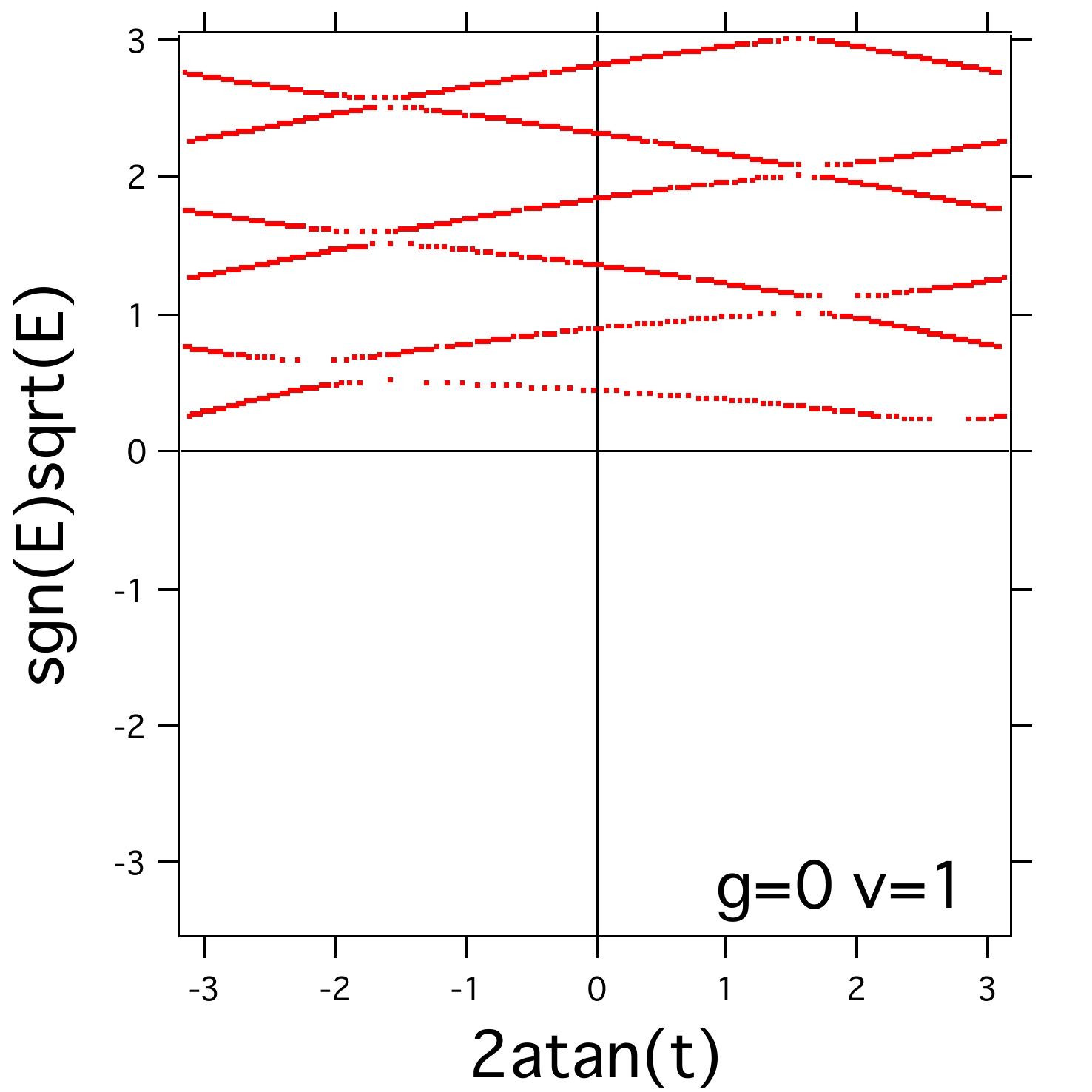}
\end{center}
\caption{energy levels of linear system fixed $g=0$ and for each $v$.}
\label{fig:two}
\end{figure}

%
\subsection{Energy levels of nonlinear system as functions of connection condition parameters}
The energy levels as functions of connection parameters $t$ and $v$ are calculated for non-zero nonlinearity parameter.  The result for $g=5$ is shown in Figure 3, while one for $g=-5$ is shown in Figure 4.  Scaling conventions are the same as in the linear case, Figure 2.  
The energy eigenvalues were plotted on the $t-E$ plane for $\delta$-strength $v=-1$ (left figure), $0$ (middle figure), $1$ (right figure) .
\begin{figure}[htbp]
\begin{center}
\includegraphics[width=4.1cm]{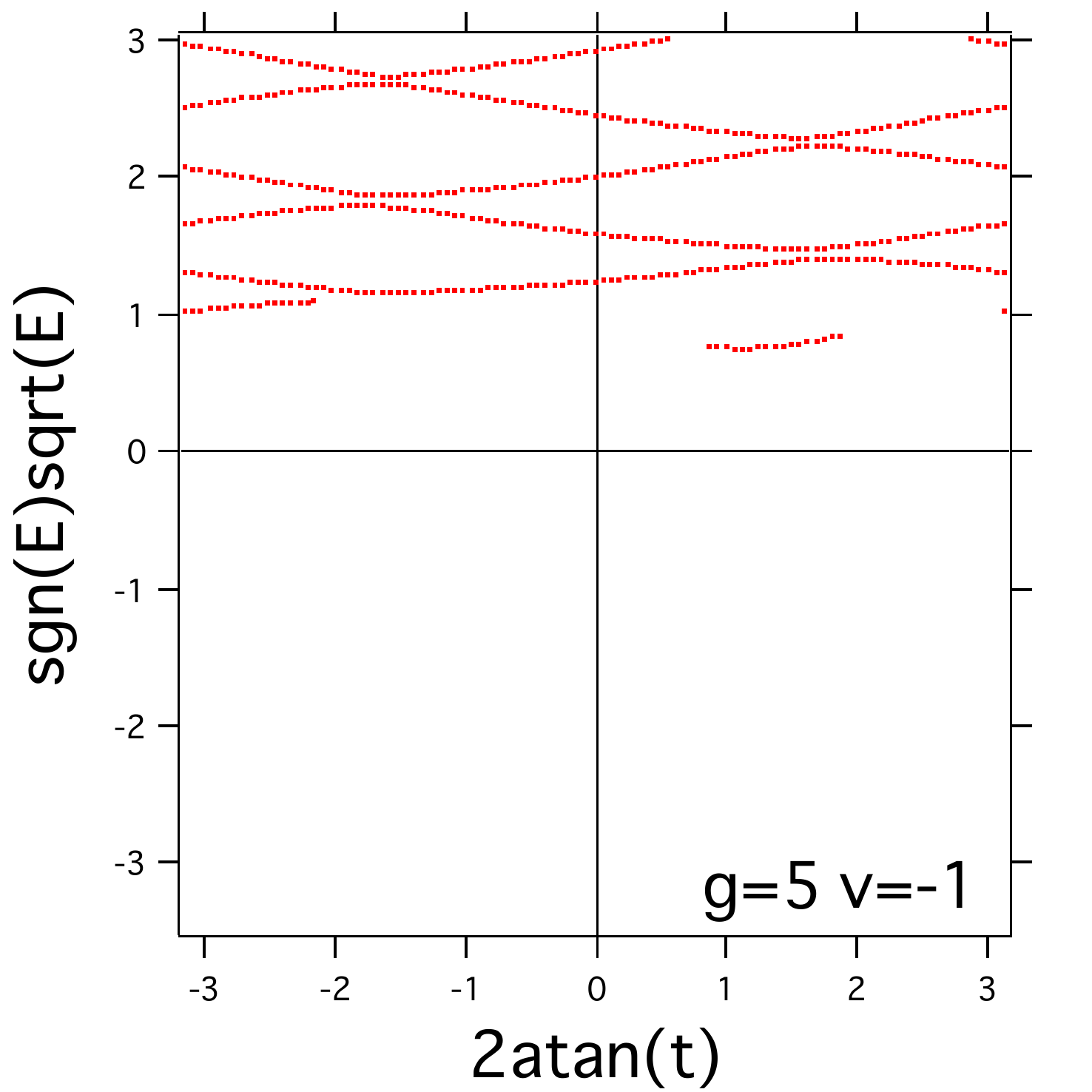}
\includegraphics[width=4.1cm]{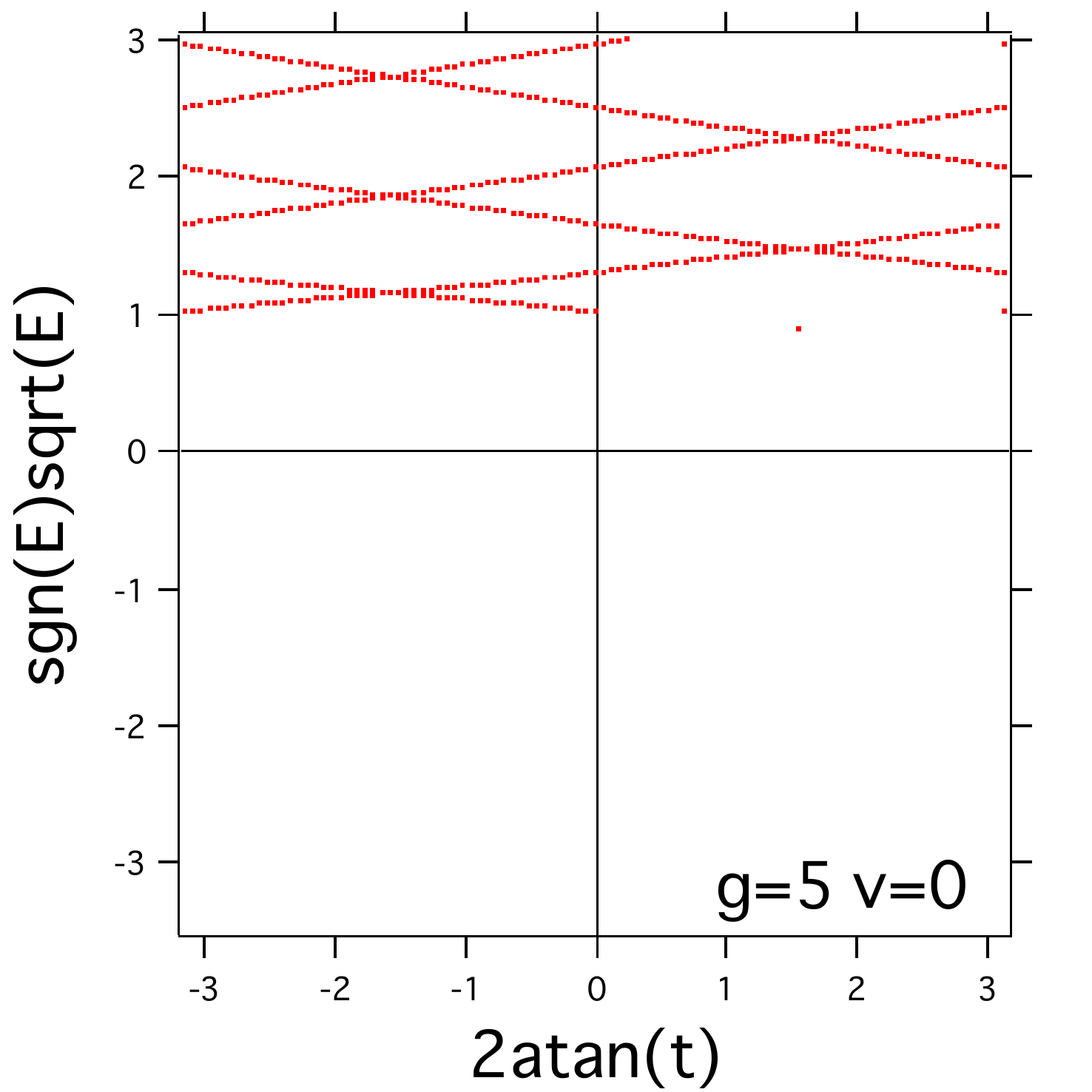}
\includegraphics[width=4.1cm]{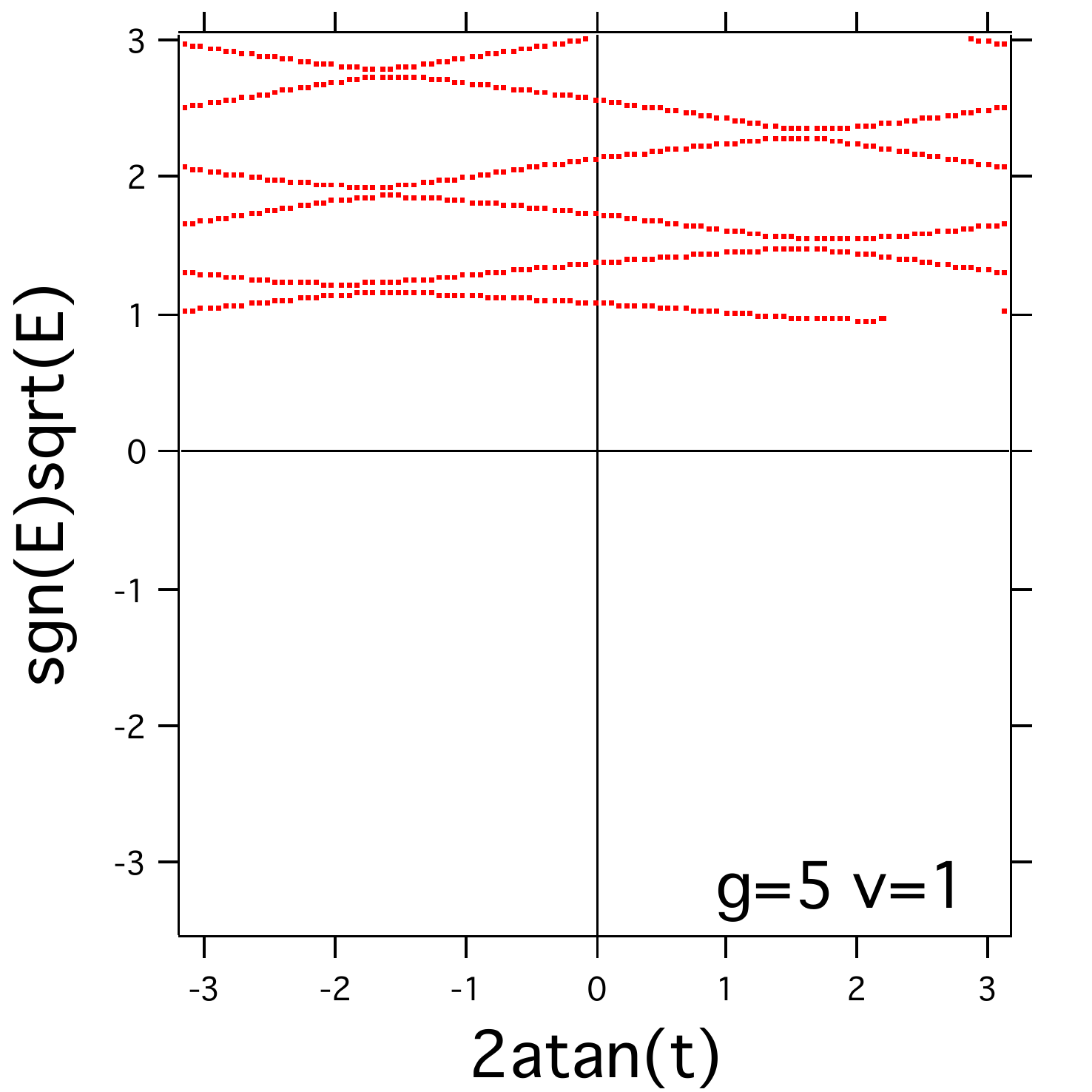}
\end{center}
\caption{The energy levels of nonlinear system, $g=5$, with varying $t$ and several values of $v$. Note the isolated points and lines around $t\sim1 \& t\sim\infty $in graphs.}
\label{fig:three}
\end{figure}

For $g=5$, Figure 3, the energy levels get deformed and are generally pushed upward, compared to the linear case, Figure 2.  All solutions are in the region $E>0$ and thus obtained from (\ref{ee21}), (\ref{ee22}) and (\ref{ee23}).  The existence of the level crossing at $\{t, v\} = \{ 1, 0\}$ and $\{ -1, 0\}$ remains intact, along with the avoided crossing around those points in parameter space $\{t, v\}$.  A very notable difference in $g=5$ case is the sudden disappearance  of energy levels. 
Eigenvalues that disappear are those calculated from (\ref{ee21}), and beyond the value of $t$ that corresponds to the disappearing point,  the condition $c<\frac{E^2}{4g}$ is violated and no solution exists. 

For $g=-5$, Figure 4, the energy levels get deformed and are generally pushed downward, compared to the linear case, Figure 2.  For this case also, the existence of the level crossing at $\{t, v\} = \{ 1, 0\}$ and $\{ -1, 0\}$ remains intact, along with the avoided crossing around those points in parameter space $\{t, v\}$.  

Unlike the case of positive $g$, we have, for $g=-5$, no disappearance of the energy level is observed.  Instead, we observe a characteristic closed-ring shaped energy level in $t-E$ plane.  The ring-shaped levels correspond to the branched-out lines in Figure 1, that are obtained from the solutions  (\ref{ee29}), while the normal energy levels are obtained from the solutions  (\ref{ee27}) and (\ref{ee28}).
In the three-dimensional plane $t-v-E$, these ring-shaped levels form a foam like structure, which has contact to the normal energy surface at the single point that lie on the line $v=0$.
Finally, it is also noteworthy that the lowest energy level appears almost symmetrically with respect to the reflection at $t=0$ axis. 
\begin{figure}[htbp]
\begin{center}
\includegraphics[width=4.1cm]{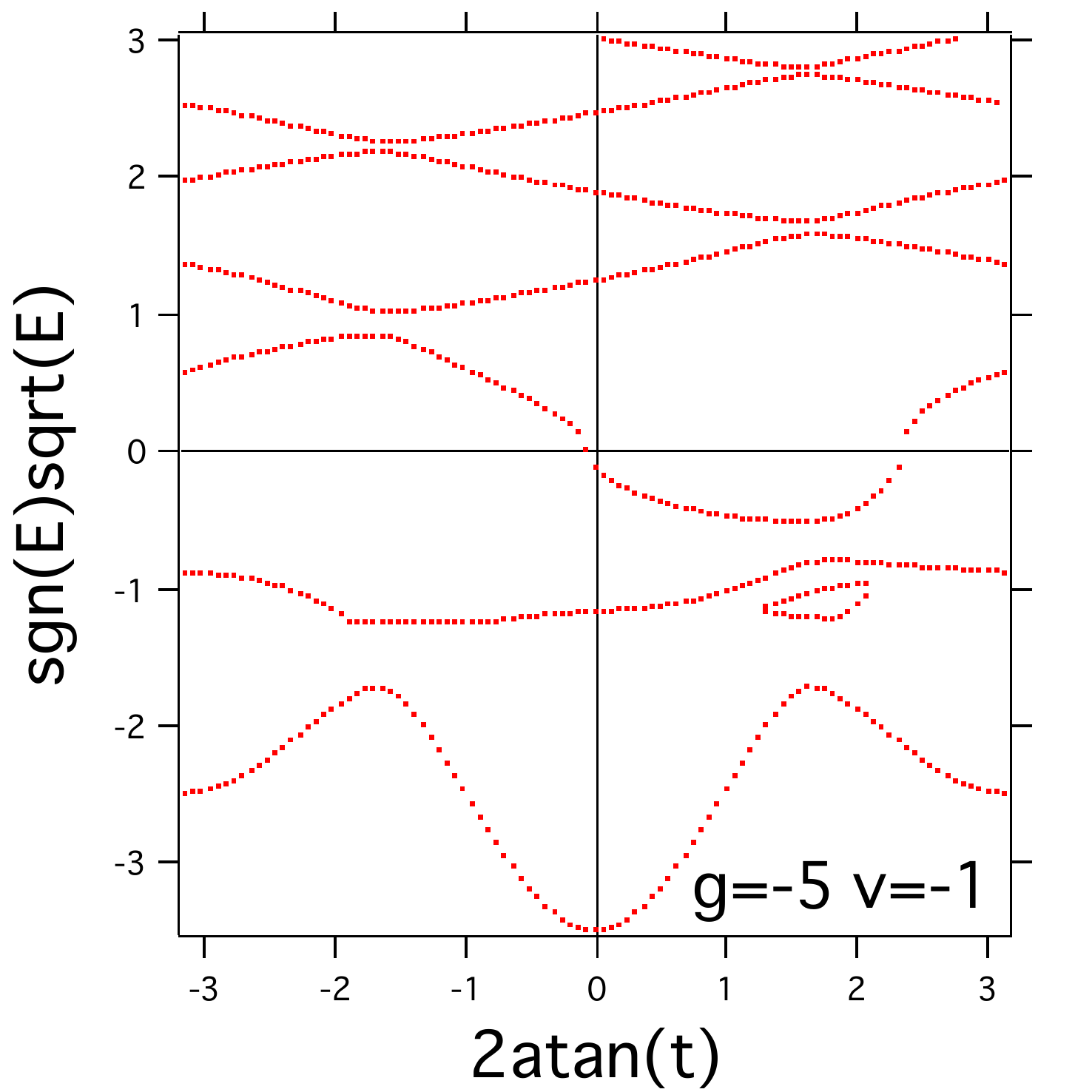}
\includegraphics[width=4.1cm]{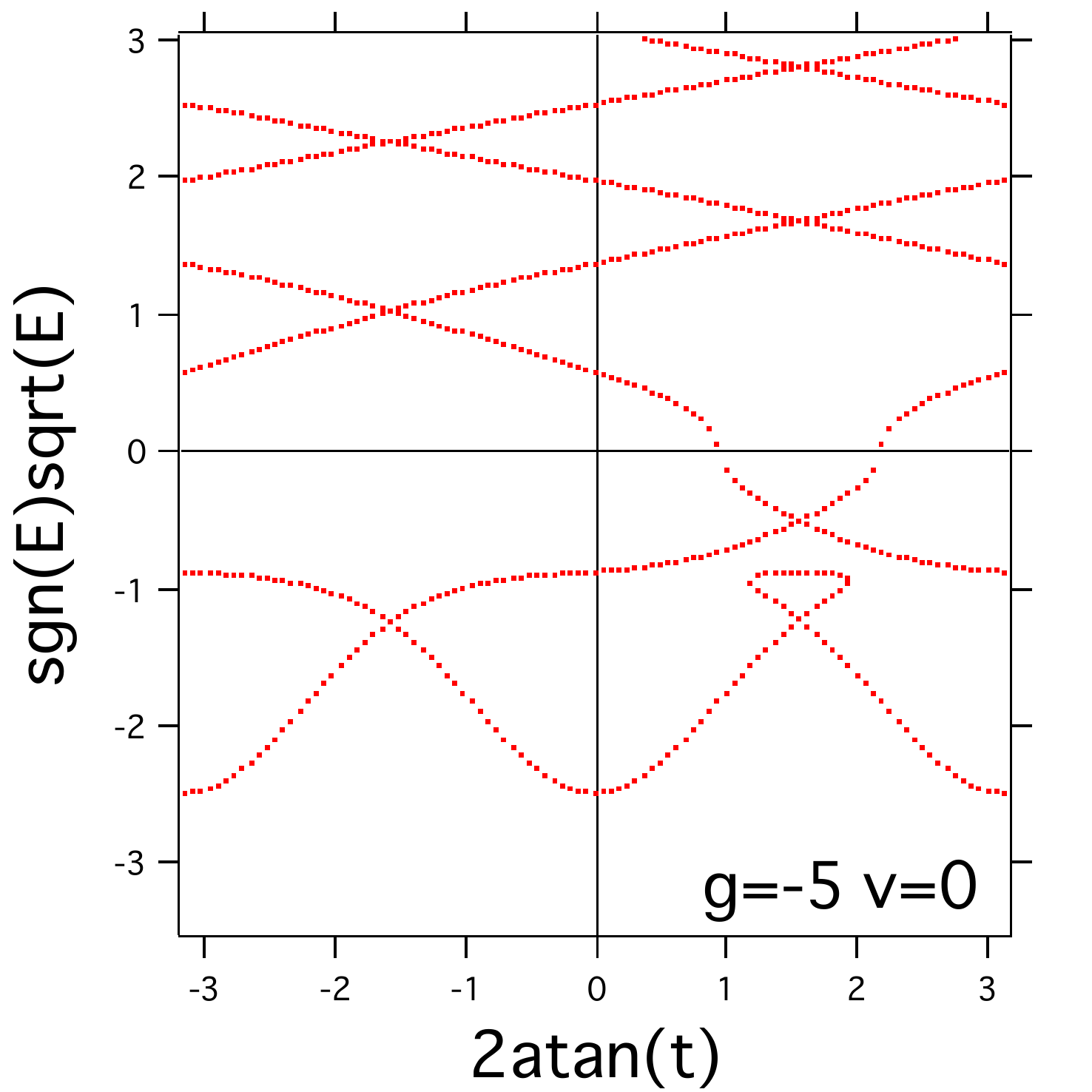}
\includegraphics[width=4.1cm]{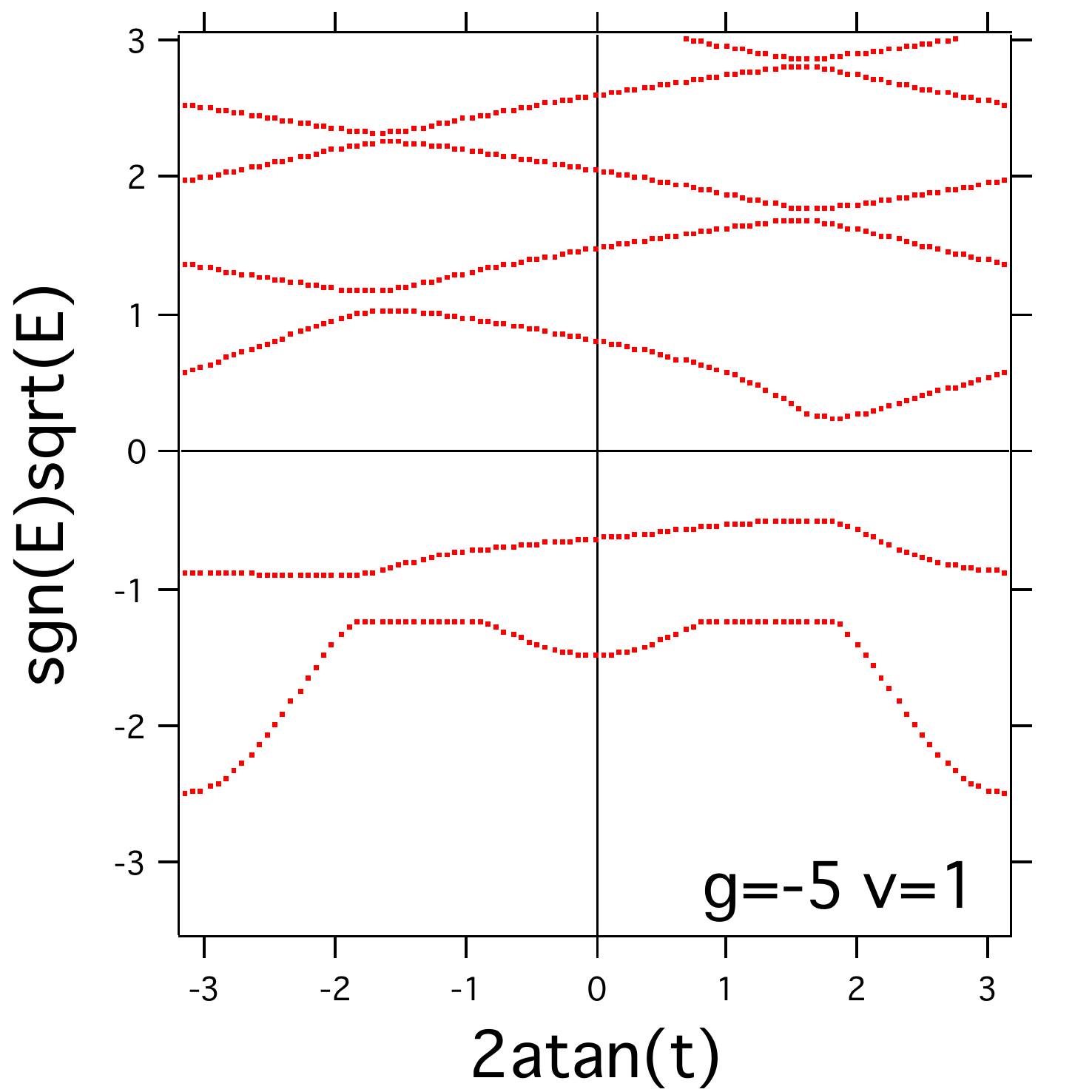}
\caption{The energy levels of nonlinear system, $g=-5$, with varying $t$ and several values of $v$.}
\label{fig:four}
\end{center}
\end{figure}
%

%
\section{Quantum holonomy}
\subsection{Berry phase}
Consider a closed path in the parameter space $\{ v, t\}$ that goes around the point where energy is degenerate. Berry has proven that, for the system giverned by Schr{\"o}dinger equation, an extra nontrivial phase might appear for the wave functions of the system, when the parameter is adiabatically changed along the path and brought back to the original value after the circulation around the degeneracy point \cite{BE84}.
It is interesting to check whether analogous phenomenon is observed in the system governed by nonlinear Schr{\"o}dinger equation. 
\begin{figure}[h]
\begin{center}
\includegraphics[width=7.5cm]{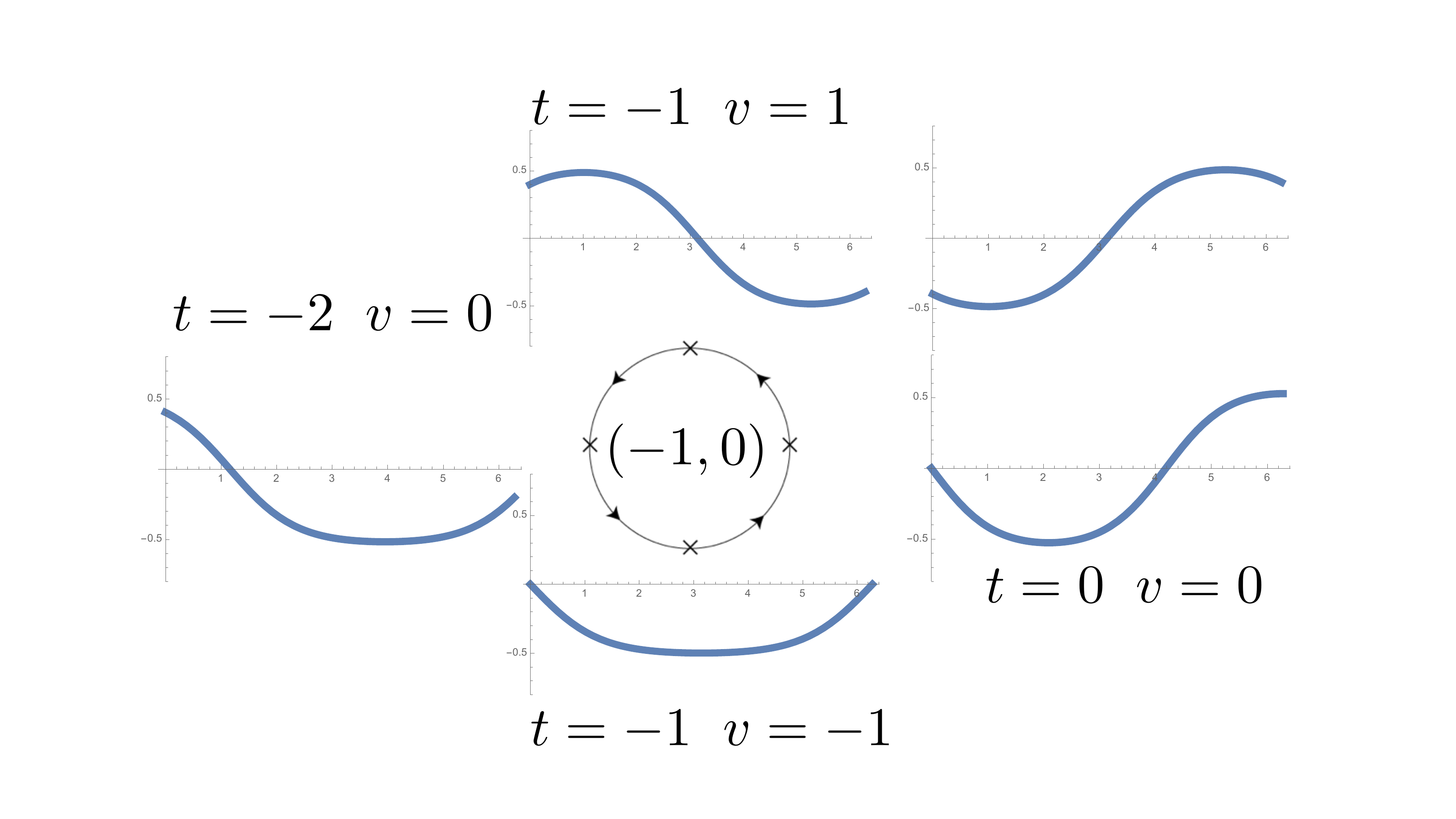}
\end{center}
\caption{Wave functions at various parameter values around $t=-1$, $v=0$ for the nonlinearity $g=-5$, showing the existence of the Berry phase $e^{i \pi}$}
\label{fig:six}
\end{figure}
\begin{figure}[h]
\begin{center}
\includegraphics[width=7.5cm]{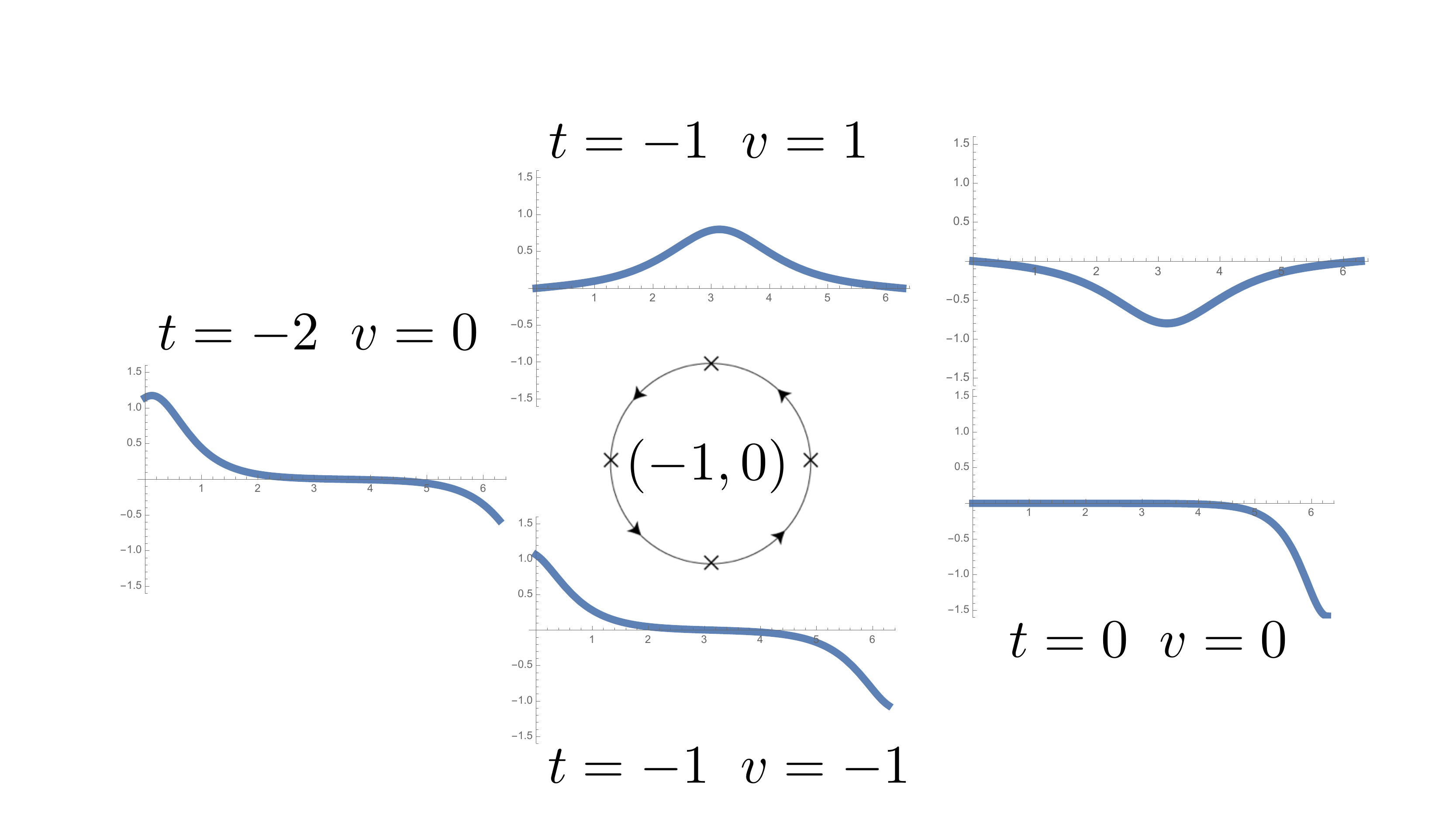}
\end{center}
\caption{Wave functions at various parameter values around $t=-1$, $v=0$ for the nonlinearity $g=5$, showing the existence of the Berry phase $e^{i \pi}$}
\label{fig:six}
\end{figure}

We can be assured that the answer is positive, from the Figure 5 and  Figure 6, in which adiabatic change of wave functions are drawn around the degeneracy points $\{ t, v \}=\{ 1, 0\}$ and $\{ t, v \}=\{ -1, 0\}$, respectively. 
It is to be noted that no Berry phase is observed for the path going around the degeneracy point connecting the ring-shaped  levels and normal energy level.

%
\subsection{Exotic quantum holonomy}
%
It is known that an adiabatic motion along a parametric cycle can bring not only the Berry phase, but also the exchange of different energy eigenvalues.  This phenomenon is known as the exotic quantum holonomy \cite{CT09}. 

In figure 7, numerically calculated eigenvalues of nonlinear Schr{\"o}dinger equation are drawn as functions of $\delta$-strength parameter $v$ with fixed value for the scale-invariant parameter $t=1$, for fifferent nonlinearity parameters.  The conditions $v=\infty$ and $v=- \infty$ are essentially equivalent, signifying the disconnected Dirichlet boundary conditions at the defect.  Therefore we can identify the left and right edges of each graphs in Figure 7, and regard the motion from the left edge, $v=-\infty$ to right edge $v=\infty$ as a cycle.

We can then recognize the existence of exotic quantum holonomy in each graphs.  In all three systems with different nonlinearity parameters $g$, the ground state and fist, third and fifth excited states gets the energy shift after the completion of adiabatic cycle $v=-\infty$ $\to$ $v=\infty$.    
\begin{figure}[htbp]
\begin{center}
\includegraphics[width=4.1cm]{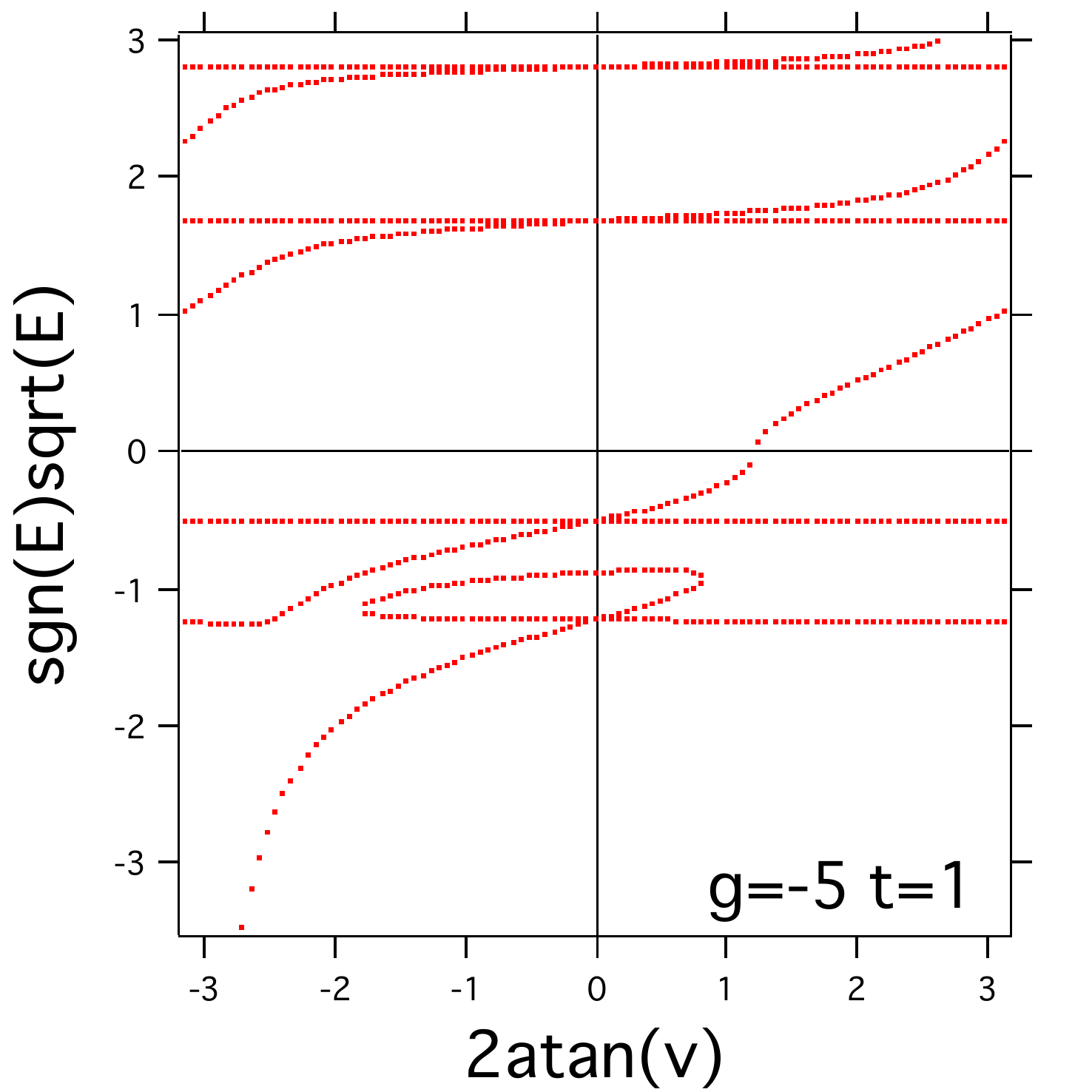}
\includegraphics[width=4.1cm]{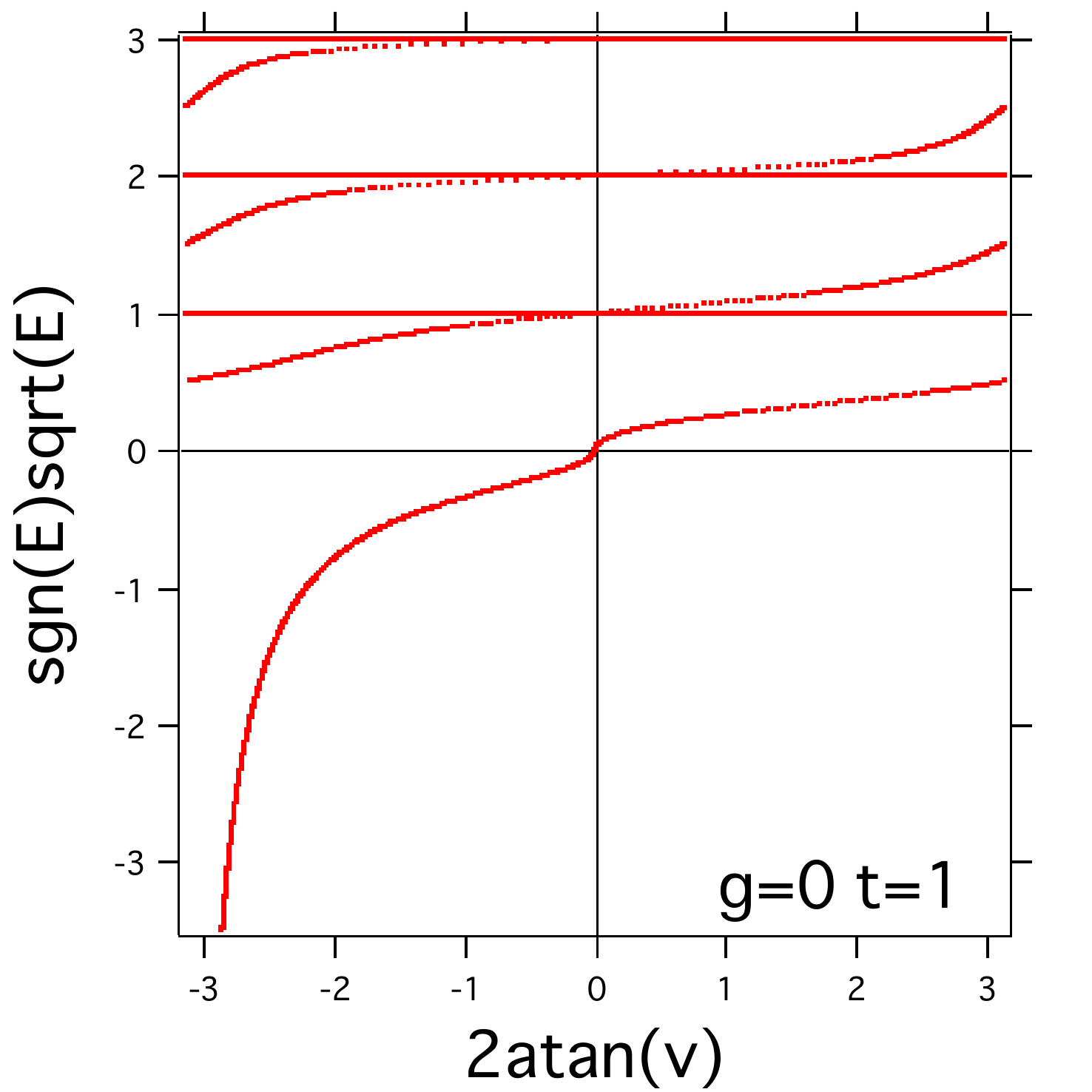}
\includegraphics[width=4.1cm]{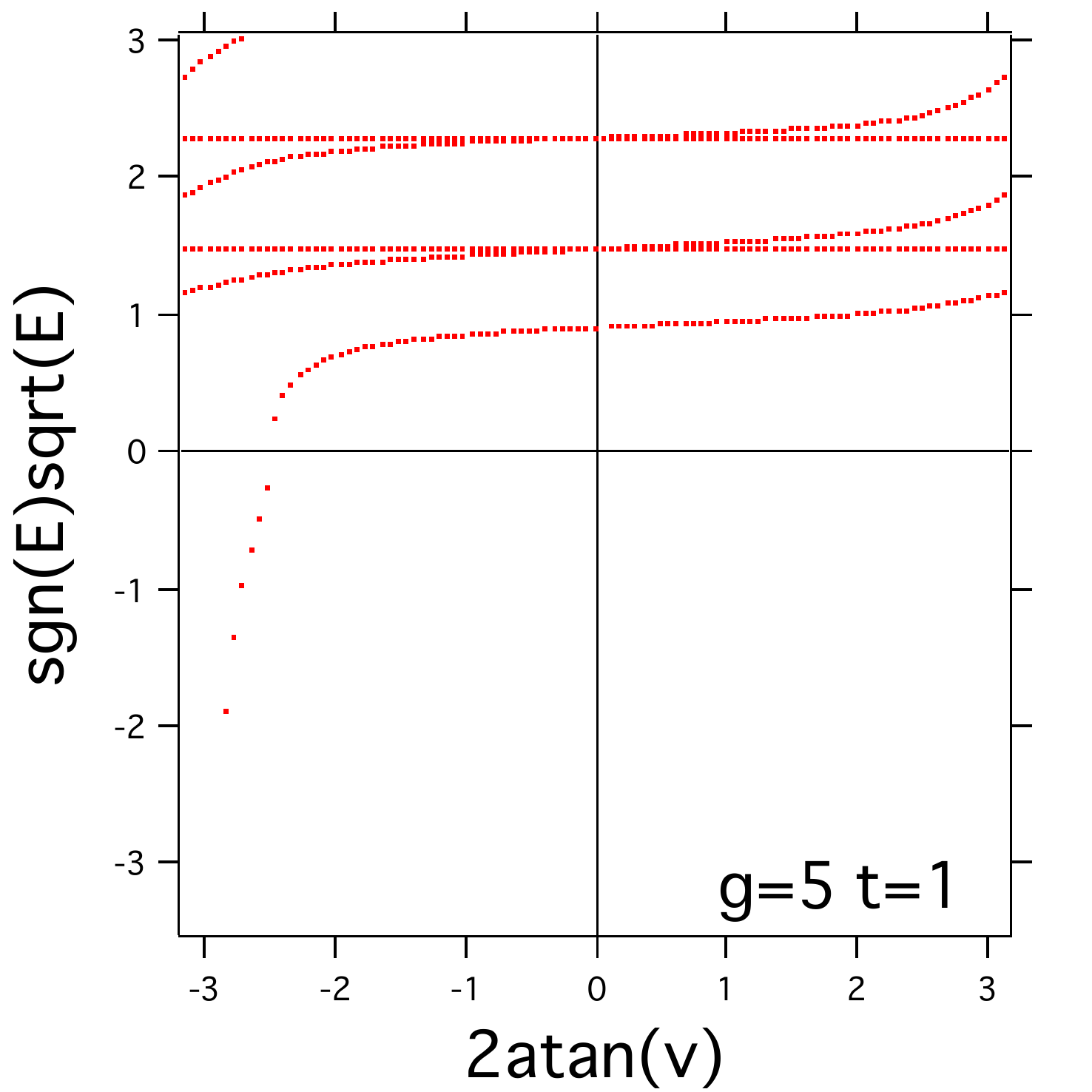}
\end{center}
\caption{Energy levels of nonlinear system fixed $t=1$ and for each $g$ .}
\label{fig:five}
\end{figure}

Thus, we have shown that exotic holonomy exists in nonlinear Schr{\"o}dinger system as well as in normal linear Schr{\"o}dinger system.
%

%
\section{Summary}

In this article, we have numerically analyzed the energy eigenvalues of a system described by cubic nonlinear Schr{\"o}dinger equation with coupling $g$ on a ring with a defect, which is described by two parameters $v$, the $\delta$-strength, and $t$, the scale invariant Fulop-Tsutsui coupling.   We have found that, for all values of nonlinear coupling parameter $g$, the degeneracy occurs at the points $\{t, v\}=\{1, 0\}$ and $\{t, v\}=\{-1, 0\}$.  Around thedegeneracy points, we have confirmed the existence of Berry phase $e^{i \pi}$.  We have also found that there is a exotic quantum holonomy expressed as the level flow as we increase $v$ and pass though $v=\infty$ and come back to finite $v$ from negative infinity $v=-\infty$.
We have also identified, in our system, the disappearing energy levels and the foam like energy surface in parametric space, the phenomena that are characteristic to nonlinear systems.

\bigskip\bigskip
\noindent{\bf Acknowledgements}

This research was supported by the Japan Ministry of Education, Culture, Sports, Science and Technology under the Grant number 15K05216.
We thank Dr. Riccardo Adami, Dr. Atushi Tanaka, and Dr. Ondrej Turek for stimulating discussions.

\bigskip

\end{document}